\documentclass{article}
\usepackage{amssymb}
\usepackage{amsfonts}
\usepackage{amsmath}

\newtheorem{theorem}{Theorem}

\newtheorem{proposition}[theorem]{Proposition}
\newtheorem{remark}[theorem]{Remark}

\newenvironment{proof}[1][Proof]{\noindent\textbf{#1.} }{\ \rule{0.5em}{0.5em}}
\input{tcilatex}

\begin{document}

\title{Tidal tensors in the description of gravity and electromagnetism}
\author{Nicoleta VOICU \\
"Transilvania" University, Brasov, Romania}
\maketitle

\begin{abstract}
Abstract: In 2008-2009, F. Costa and C. Herdeiro proposed a new gravito -
electromagnetic analogy, based on tidal tensors. We show that connections on
the tangent bundle of the space-time manifold can help not only in finding a
covnenient geometrization of their ideas, but also a common mathematical
description of the main equations of gravity and electromagnetism.
\end{abstract}



\section{Introduction}

In two recent papers,\cite{Costa}, \cite{Costa2}, F. Costa and C. Herdeiro
provided a new gravito-magnetic analogy, meant to overcome the limitations
of the two classical ones (namely, the linearized approach, which is only
valid in the case of a weak gravitational field and the one based on Weyl
tensors, which compares tensors of different ranks). The central role in
this analogy is played by worldline deviation\ equations and the resulting
tidal tensors; it is in terms of these tensors that the fundamental
equations of the gravitational and electromagnetic fields are expressed and
compared. We argue that this is not only a natural idea, but an idea which
can be brought further -- it can underlie more than just an analogy between
the two fields: a common geometric model for these.

Still, in the cited papers, in order to be able to make such an analogy, it
is imposed a restriction: in the case of worldline deviation for charged
particles in flat Minkowski space, covariant derivatives of the deviation
vector field $w$ are required to identically vanish along the initial
worldline.

We have shown in a previous paper, \cite{Voicu1}, that, if we raise to the
tangent bundle $TM$ of the space-time manifold and use an appropriate
1-parameter family of Ehresmann connections $\overset{\alpha }{N}$, we have
at least two advantages. On one side, there is no longer need of any
restriction upon $w$ -- the "work" of eliminating the unwanted term in the
worldline deviation equation is taken by the adapted frame. On the other
side, the obtained tidal tensor expressions for the basic equations of the
two physical fields are valid not only in the case when we have either
gravity only, or electromagnetic field alone (as in \cite{Costa}, \cite%
{Costa2}), but also in the general case, when both are present -- thus, they
also provide a common geometric language for the two physical fields.

The Ehresmann connections we defined in \cite{Voicu1} give rise to very
convenient frames on $TM:$ adapted frames. As for covariant derivatives of
tensors, we will use the ones given by some special affine connections $%
\overset{\alpha }{D}$ on the tangent bundle. In \cite{Voicu1}, we defined a
first variant of such a 1-parameter family of connections. In the present
paper, we propose a different choice for $\overset{\alpha }{D}$, with richer
properties:\ 1)\ lifts of worldlines of charged particles are autoparallel
curves for $\overset{\alpha }{D}$; 2)\ geodesic deviation equations are as
simple as possible; 3) Riemann (and Ricci)\ tensors of $\overset{\alpha }{D}$
can be obtained from tidal tensors just by differentiating the latter with
respect to the fiber coordinates on $TM$. Also, we discuss in more detail
the obtained equations.

The idea we use here -- of encoding gravity in the metric and
electromagnetism, in Ehresmann connections (together with affine
connections) on the tangent bundle\footnote{%
Other attempts of unifying gravity and electromagnetism, based on tangent
bundle geometry, try to include information regarding electromagnetism in
Finsler-type metrics (Randers, Beil or Weyl metrics, \cite{Beil}, \cite%
{Beil2}). Also, recently, Wanas, Youssef and Sid-Ahmed produced another
description, \cite{Wanas}, based on teleparallelism on $TM$. Still, we
adopted Miron's idea as leading to relatively simple computations and to
elegant expressions for Einstein and Maxwell equations.} $TM$ -- was
proposed by R. Miron and collaborators, \cite{Miron-Buchner}, \cite{Ingarden}%
, \cite{Lagrange}; just, here we use different connections, meant to offer a
more convenient expression for worldline deviation equations.

The paper is organized as follows. In Section 2, we present the elements of
the gravito-electromagnetic analogy by Costa and Herdeiro (\cite{Costa}, 
\cite{Costa2}) which are necessary in the subsequent. In the following three
sections, we introduce the Ehresmann connections $\overset{\alpha }{N},$ the
affine connections $\overset{\alpha }{D}$ and study geodesics, together with
geodesic deviation. Section 6 is devoted to the geometric expressions of the
basic equations of the two physical fields. In the last section, we rewrite
in terms of adapted derivatives Costa and Herdeiro's equations and point out
that the restriction imposed in (\cite{Costa}, \cite{Costa2}) upon the
deviation vector field is no longer necessary.

\section{Tidal tensors and gravito-electromagnetic a\-na\-lo\-gy}

Consider a 4-dimensional Lorentzian manifold $(M,g),$ with signature $%
(-,+,+,+),$ regarded as space-time manifold, with local coordinates $%
(x^{i})_{i=\overline{0,3}}$ and Levi-Civita connection $\nabla .$ Throughout
the paper, we will mean by $(\partial _{i})$ the natural basis of the module
of vector fields on $M;$ the speed of light $c$ and the gravitational
constant $k$ will be considered as equal to 1.

Worldlines of particles subject to \textit{gravity only} are geodesics $%
s\mapsto (x^{i}(s))$\ of $(M,g):$

\begin{equation*}
\dfrac{\nabla u^{i}}{ds}=0,~\ \ \ u=\dot{x},
\end{equation*}%
where $s$ is the natural parameter (i.e., $g_{ij}u^{i}u^{j}=-1$). Curvature
of space-time becomes manifest in the \textit{geodesic deviation equation}: 
\begin{equation}
\dfrac{\nabla ^{2}w^{i}}{ds^{2}}=e_{~k}^{i}w^{k},~\ \ \ \ \ \
e_{~k}^{i}=r_{j~kl}^{~i}u^{j}u^{l},  \label{classical_Jacobi}
\end{equation}%
where $w=w^{i}\partial _{i}$ is the deviation vector field and $e_{~k}^{i}$
define the so-called \textit{tidal (electrogravitic)\ tensor\footnote{%
Here, we have used a different sign convention for the Riemann tensor ($%
r_{j~kl}^{~i}=\partial _{l}\gamma _{jk}^{i}-\partial _{k}\gamma
_{jl}^{i}+\gamma _{jk}^{h}\gamma _{hl}^{i}-\gamma _{jl}^{h}\gamma _{hk}^{i}$%
) than in \cite{Costa}, \cite{Costa2}, resulting in a different sign for $%
e_{~j}^{i}.$}.}

On the other side, in special relativity\textit{\ }(where $%
g_{ij}=diag(-1,1,1,1)$),\textit{\ }the electromagnetic field is described by
the 4-\textit{potential }1-form $A=A_{i}(x)dx^{i}$ and the electromagnetic
2-form $F=dA,$ i.e.,%
\begin{equation}
F=\dfrac{1}{2}F_{ij}dx^{i}\wedge dx^{j},~\ \ F_{ij}=\partial
_{i}A_{j}-\partial _{j}A_{i}.  \label{comps_F_A}
\end{equation}

Worldlines of charged particles subject to an\textit{\ }electromagnetic field%
\textit{\ }are solutions of the Lorentz equations:%
\begin{equation}
\dfrac{\nabla u^{i}}{ds}=\dfrac{q}{m}F_{~j}^{i}u^{j},~\ \ u=\dot{x};
\label{Lorentz_eq_flat}
\end{equation}%
here,$\ \dfrac{\nabla u^{i}}{ds}=\dfrac{du^{i}}{ds},$ $q$ is the electric
charge of the particle and $m,$ its mass. For families of worldlines of
particles with\ same ratio $\dfrac{q}{m},$ one can determine the \textit{%
worldline deviation equation:}%
\begin{equation}
\dfrac{\nabla ^{2}w^{i}}{ds^{2}}=\dfrac{q}{m}(E_{~k}^{i}w^{k}+F_{~k}^{i}%
\dfrac{\nabla w^{k}}{ds}),  \label{worldline_deviation_classic}
\end{equation}%
where 
\begin{equation}
E_{~k}^{i}=u^{j}\nabla _{\partial _{k}}F_{~j}^{i}.
\label{electric_tidal_tensor}
\end{equation}

\bigskip

Traditionally, \cite{Costa}, \cite{worldline-deviation}, it is imposed the
restriction that, along the initial worldline:%
\begin{equation}
\dfrac{\nabla w^{i}}{ds}=0;\   \label{**}
\end{equation}%
under this assumption, the worldline deviation equations reduce to: 
\begin{equation}
\dfrac{\nabla ^{2}w^{i}}{ds^{2}}=\dfrac{q}{m}E_{~k}^{i}w^{k},
\label{simplified_em_dev}
\end{equation}%
which makes it possible to compare (\ref{classical_Jacobi}) and (\ref%
{worldline_deviation_classic}). Following the analogy $E_{~k}^{i}~\sim
e_{~k}^{i},$ \cite{Costa}, \cite{Costa2}, Maxwell's equations\textit{\ }are
written (after contracting with the 4-velocity $u$) as:%
\begin{equation}
\ 
\begin{array}{l}
\nabla _{\partial _{i}}F^{ij}=4\pi J^{i} \\ 
\nabla _{\partial _{i}}F_{jk}+\nabla _{\partial _{k}}F_{ij}+\nabla
_{\partial _{j}}F_{ki}=0%
\end{array}%
~~\ \ \ \ 
\begin{array}{c}
\Rightarrow  \\ 
\Rightarrow 
\end{array}%
~\ \ \ \ \ \ 
\begin{array}{l}
E_{~i}^{i}=-4\pi \rho _{c} \\ 
E_{[ij]}=\dfrac{1}{2}u^{k}\nabla _{\partial _{k}}F_{ij},%
\end{array}
\label{Costa_Maxwell}
\end{equation}%
(where square brackets denote antisymmetrization and $\rho _{c}=-J_{i}u^{i}$%
), while gravitational field equations\textit{\ }take the form:%
\begin{equation}
\begin{array}{l}
r_{ij}=8\pi (T_{ij}-\dfrac{1}{2}g_{ij}T_{~l}^{l}) \\ 
r_{jk}=r_{kj}%
\end{array}%
~\ \ \ \ \ \ 
\begin{array}{c}
\Rightarrow  \\ 
\Rightarrow 
\end{array}%
~\ \ \ \ \ 
\begin{array}{l}
e_{~i}^{i}=-4\pi (2\rho _{m}-T_{~i}^{i}) \\ 
e_{[ij]}=0,%
\end{array}
\label{Costa_gravi}
\end{equation}%
where $T_{ij}$ is the stress-energy tensor and $\rho _{m}=T_{ij}u^{i}u^{j}$.

\begin{remark}
Passing to the tangent bundle, we will be able to regard velocities as fiber
coordinates and contract with respect to the latter in defining tidal
tensors. This way, all the four implications (\ref{Costa_Maxwell}), (\ref%
{Costa_gravi}) become equivalences (we can "discard"\ the fiber coordinates
from the tidal tensor equations by differentiating with respect to them,
thus getting again the classical versions of the equations).
\end{remark}

In the cited papers, the authors also use the analogues of the above tidal
tensors, built from the Hodge duals of the 2-forms $r$ and $F,$ resulting in
two more pairs of equations (\ref{Costa_Maxwell}), (\ref{Costa_gravi}).
Taking into account the above remark, (\ref{Costa_Maxwell}) and (\ref%
{Costa_gravi}) are sufficient for our purposes, so we have omitted these two
extra pairs of equations.

\section{Ehresmann connections}

Consider now the tangent bundle $(TM,\pi ,M)$ of the 4-dimensional
Lorentzian manifold $(M,g)$; here, we denote the local coordinates by $%
(x\circ \pi ,y)=:(x^{i},y^{i})_{i=\overline{0,3}}$ and by $_{,i}$ and $\cdot
_{i},$ partial differentiation with respect to $x^{i}$ and $y^{i}$
respectively. An (arbitrary) Ehresmann connection $N$ on $TM,$ \cite%
{Lagrange}, \cite{Shen}, gives rise to the adapted basis 
\begin{equation}
(\delta _{i}=\dfrac{\partial }{\partial x^{i}}-N_{~i}^{j}(x,y)\dfrac{%
\partial }{\partial y^{j}},~~\ \ \dot{\delta}_{i}=\dfrac{\partial }{\partial
y^{i}}),  \label{general_adapted_basis}
\end{equation}%
and to its dual $(dx^{i},\delta y^{i}=dy^{i}+N_{~j}^{i}dx^{j}).$ For a
vector field $X=X^{i}\delta _{i}+\tilde{X}^{i}\dot{\delta}_{i},$ we denote
by $hX=X^{i}\delta _{i}$ and $vX=\tilde{X}^{i}\dot{\delta}_{i}$ its
horizontal and vertical components respectively.

The horizontal 1-form $l=l_{i}dx^{i},$ where 
\begin{equation}
l^{i}=\dfrac{y^{i}}{\left\Vert y\right\Vert },~~\left\Vert y\right\Vert =%
\sqrt{g_{ij}y^{i}y^{j}},  \label{supporting element}
\end{equation}%
(and indices of $l$ are lowered by means of $g_{ij}$) is called the
distinguished section on $TM$, \cite{Shen},

A curve $c:t\mapsto (x^{i}(t))$ on $M$ (where $t$ is an arbitrary parameter)
is an autoparallel curve\textit{\ (geodesic}) for $N$ if: 
\begin{equation}
\dfrac{\delta y^{i}}{dt}=\dfrac{dy^{i}}{dt}+N_{~j}^{i}(x,y)y^{j}=0,~\ \ \ \
y^{i}=\dfrac{dx^{i}}{dt};  \label{autoparallel_N}
\end{equation}

By direct computation, it follows that deviations of geodesics (\ref%
{autoparallel_N}) are described by: 
\begin{equation}
\dfrac{\delta ^{2}w^{i}}{dt^{2}}=R_{~jk}^{i}(x,y)y^{k}w^{j}+\mathbb{T}%
_{~j}^{i}(x,y)\dfrac{\delta w^{j}}{dt},  \label{deviations_N}
\end{equation}%
where $w^{i}$ are the components of the deviation vector field (on $M$),%
\begin{equation}
R_{jk}^{i}=\delta _{k}N_{~j}^{i}-\delta _{j}N_{~k}^{i}  \label{curvature_N}
\end{equation}%
are the components of the$~$curvature of\textit{\ }$N$ and 
\begin{equation*}
\mathbb{T}_{~j}^{i}=y^{k}N_{~k\cdot j}^{i}-N_{~j}^{i},
\end{equation*}%
those of the~\textit{strong torsion }of $N$, \cite{Miron-Anastasiei}.

\bigskip

\textbf{Particular case. }A\ \textit{spray connection, \cite{Anto},} is an
Ehresmann connection with the property that there exist some real-valued
functions $G^{i}=G^{i}(x,y)$ with the properties:

\begin{enumerate}
\item $N_{~j}^{i}=G_{~\cdot j}^{i}=:G_{~j}^{i};$

\item $G^{i}(x,\lambda y)=\lambda ^{2}G^{i}(x,y),$ $\forall \lambda \in 
\mathbb{R}$;

\item with respect to coordinate changes on $TM,$ the functions $G^{i}$
behave in such a way that the \textit{spray} $S=y^{i}\dfrac{\partial }{%
\partial x^{i}}-2G^{i}\dfrac{\partial }{\partial y^{i}}$ is a vector field.
\end{enumerate}

For spray connections, the strong torsion $\mathbb{T}$ identically vanishes.
Conversely, if for a connection $N,$ the strong torsion is identically 0,
then, by setting, $2G^{i}:=N_{~k}^{i}y^{k},$ it follows that $2G_{~\cdot
j}^{i}=N_{~k\cdot j}^{i}y^{k}+N_{~j}^{i}=2N_{~j}^{i},$ hence, $N$ is a spray
connection.

Consequently: spray connections are the only Ehresmann connections with the
property that deviations of autoparallel curves are described by equations
whose right hand side does not depend on the derivatives of $w:$%
\begin{equation}
\dfrac{\delta ^{2}w^{i}}{dt^{2}}=E_{~j}^{i}w^{j},  \label{deviations_N_tidal}
\end{equation}%
where $E_{~j}^{i}=R_{~jk}^{i}y^{k}.$

We will call the quantity%
\begin{equation}
E~\mathbb{=~}E_{~j}^{i}\delta _{i}\otimes dx^{j},~\ ~\ \ \
E_{~j}^{i}=~R_{~jk}^{i}y^{k};  \label{tidal_tensor}
\end{equation}%
the \textit{tidal tensor }of $N;$ for spray connections, this tensor
contains all the information regarding deviations of autoparallel curves.

\begin{remark}
Geodesic equations for a spray connection can be also written, \cite{Anto}, 
\cite{Lagrange}, as: 
\begin{equation}
\dfrac{dy^{i}}{dt}+2G^{i}(x,y)=0,~\ y=\dot{x}.  \label{spray_geodesics}
\end{equation}%
Conversely, if equations (\ref{spray_geodesics})\ are globally defined and
the functions $G^{i}$ are 2-homogeneous in $y,$ then, \cite{Bucataru}, $%
G^{i} $ are the coefficients of a spray on $TM.$
\end{remark}

\bigskip

As an application of the above, consider the following 1-parameter family of 
\textit{Randers-type Finslerian fundamental functions} \cite{Shen}, \cite%
{Ingarden} depending on $\alpha :$%
\begin{equation}
\overset{\alpha }{L}~=\sqrt{g_{ij}(x)\dot{x}^{i}\dot{x}^{j}}+\alpha A_{i}%
\dot{x}^{i};  \label{randers}
\end{equation}%
here, $g_{ij}$ is the Lorentzian metric on $M$ as above, $A=A_{i}(x)dx^{i},$
a 1-form on $M$ (momentarily, with no relation with the one in Section 2)
and $\alpha \in \mathbb{R}$ is a parameter.

Extremal curves $x=x(t)$ (i.e., $t=const\cdot s$) for the action $\int 
\overset{\alpha }{L}dt$ are given by:%
\begin{equation}
\dfrac{dy^{i}}{dt}+\gamma _{~jk}^{i}y^{j}y^{k}-\alpha \left\Vert
y\right\Vert F_{~j}^{i}y^{j}=0,~\ y^{i}=\dot{x}^{i},  \label{Lorentz_eq}
\end{equation}%
where%
\begin{equation}
F_{~j}^{i}=g^{ih}(A_{j,h}-A_{h,j}),~\ \ \ \left\Vert y\right\Vert =\sqrt{%
g_{ij}y^{i}y^{j}};  \label{F_ij}
\end{equation}%
we get a 1-parameter family of sprays on $TM,$ with coefficients $G^{i}=~%
\overset{\alpha }{G}\overset{}{^{i}}$ given by:%
\begin{equation}
2\overset{\alpha }{G}\overset{}{^{i}}(x,y)=\gamma _{~jk}^{i}y^{j}y^{k}+2%
\overset{\alpha }{B}\overset{}{^{i}},  \label{spray_randers}
\end{equation}%
where\footnote{%
Indices are raised here by means of the pseudo-Riemannian metric $g_{ij}(x).$
This feature makes our approach different from the classical treatment of
Randers geometry (where one uses a more complicated, Finsler-type, metric
tensor $g_{ij}(x,y)$, \cite{Shen}) -- and similar to the Ingarden geometry
proposed by Miron, \cite{Ingarden}; the difference between the Ehresmann
connections we build here and those in Ingarden geometry relies in the
2-homogeneity in the fiber coordinates of the term $-\alpha \left\Vert
y\right\Vert F_{~j}^{i}$ in (\ref{Lorentz_eq}), which allows us to use spray
connections.}%
\begin{equation}
2\overset{\alpha }{B}\overset{}{^{i}}=-\alpha \left\Vert y\right\Vert
F_{~j}^{i}y^{j}.  \label{definition_B}
\end{equation}

We will also use the notation: $F^{i}=F_{~j}^{i}y^{j},$ i.e., $2\overset{%
\alpha }{B}\overset{}{^{i}}:=-\alpha \left\Vert y\right\Vert F^{i}.$ If
there is no risk of confusion, we will not explicitly indicate in the
notation of spray and connection coefficients, adapted derivatives etc., the
parameter $\alpha $ (i.e., we will use $G^{i},B^{i},$ $%
G_{~j}^{i},B_{~j}^{i},...$ instead of $\overset{\alpha }{G}\overset{}{^{i}},%
\overset{\alpha }{B}\overset{}{^{i}}$ $\overset{\alpha }{G}\overset{}{%
_{~j}^{i}},\overset{\alpha }{B}\overset{}{_{~j}^{i}}$ etc.).

\bigskip

It is worth noticing some properties of the functions $B^{i}$ in (\ref%
{definition_B}).

1. Functions $B^{i},$ (\ref{definition_B}), are the components of a
horizontal vector field $B=B^{i}\delta _{i}$ on $TM.$

2. The derivatives of the functions $B^{i}$ with respect to the fiber
coordinates are:%
\begin{eqnarray}
&&B_{~j}^{i}=B_{~\cdot j}^{i}=-\dfrac{\alpha }{2}(F^{i}l_{j}+\left\Vert
y\right\Vert F_{~j}^{i})  \label{expr_Bij} \\
&&B_{~jk}^{i}:=B_{~\cdot jk}^{i}=-\dfrac{\alpha }{2}(l_{\cdot
jk}F^{i}+l_{j}F_{~k}^{i}+l_{k}F_{~j}^{i}).  \notag
\end{eqnarray}

3. From the homogeneity of degree 2 of $B$ in the fiber coordinates, it
follows: $B_{~j}^{i}y^{j}=2B^{i},~\ \ B_{~jk}^{i}y^{k}=B_{~j}^{i},~\
B_{~\cdot jkl}^{i}y^{l}=0.$

4. The spray connection coefficients of $N=~\overset{\alpha }{N}$ are
expressed in terms of $\gamma _{~jk}^{i}$ and $B$ as:%
\begin{equation}
G_{~j}^{i}=\gamma _{~jk}^{i}y^{k}+B_{~j}^{i}.  \label{spray_conn_coeff}
\end{equation}

\bigskip

\textbf{Particular case: }For \ $\alpha =0$, we get: \ $2\overset{0}{G}%
\overset{}{^{i}}=\gamma _{~jk}^{i}y^{j}y^{k},$ $\overset{0}{G}\overset{}{%
_{~j}^{i}}=\gamma _{~jk}^{i}y^{k},$ hence autoparallel curves of $\overset{0}%
{N}$ coincide with the geodesics of $(M,g).$ Moreover, we have%
\begin{equation*}
\overset{0}{R}\overset{}{_{~jk}^{i}}=r_{l~jk}^{~i}y^{l}=:r_{~jk}^{i}
\end{equation*}%
and the tidal tensor is in this case, $\overset{0}{E}\overset{}{_{~j}^{i}}%
=e_{~j}^{i}=r_{~jk}^{i}y^{k}.$

\section{Affine connections on $TM$}

In the following, we will focus on the functions $\overset{\alpha }{L}$.
Consider%
\begin{equation}
G_{~jk}^{i}:=G_{~\cdot jk}^{i}=\gamma _{~jk}^{i}+B_{~jk}^{i};
\label{Berwald_conn}
\end{equation}%
we define the affine connections $D=\overset{\alpha }{D}$ on $TM$ which act
on the $\overset{\alpha }{N}$-adapted basis vectors as: 
\begin{equation}
D_{\delta _{k}}\delta _{j}=G_{jk}^{i}\delta _{i},~\ \ D_{\delta _{k}}\dot{%
\delta}_{j}=G_{jk}^{i}\dot{\delta}_{i},~\ \ \ D_{\dot{\delta}_{k}}\delta
_{j}=D_{\dot{\delta}_{k}}\dot{\delta}_{j}=0.  \label{d-connection}
\end{equation}

\begin{remark}
Connections $\overset{\alpha }{D},$ $\alpha \in \mathbb{R},$ preserve by
parallelism the distributions generated by $\overset{\alpha }{N}$ (hence,
they are \textit{distinguished linear connections, }\cite{Lagrange}, \cite%
{Miron-Anastasiei}, on $TM$), i.e., for any two vector fields $X,Y$ on $TM,$
we have: $D_{X}(hY)=hD_{X}Y,$ $D_{X}(vY)=vD_{X}Y.$
\end{remark}

Connections $\overset{\alpha }{D}$ are generally, non-metrical.

\textbf{Particular case: }For $\alpha =0,$ we get: 
\begin{equation*}
\overset{0}{G}\overset{}{_{~jk}^{i}}=\gamma _{~jk}^{i},
\end{equation*}%
i.e., for vector fields $X,Y\ $on the base manifold $M,$ the horizontal lift 
$l_{h}(\nabla _{X}Y)$ of the Levi-Civita covariant derivative $\nabla _{X}Y$
and the $\overset{0}{D}$-covariant derivative $\overset{0}{D}%
_{l_{h}(X)}l_{h}(Y)$ of the (separately) lifted vector fields coincide: 
\begin{equation*}
l_{h}(\nabla _{X}Y)=\overset{0}{D}_{l_{h}(X)}l_{h}(Y).
\end{equation*}%
In this sense, $\overset{0}{D}$ can be considered as the $TM$-equivalent of
the Levi-Civita connection $\nabla $ and $\overset{\alpha }{D}$, as a
"perturbation" of $\overset{0}{D},$ with contortion tensor $B.$

\bigskip

Curvature and torsion tensors of $\overset{\alpha }{D}$ can be determined by
direct computation.

\begin{enumerate}
\item The torsion of $D=\overset{\alpha }{D}$ is given by:%
\begin{equation}
\mathbb{T}=R_{~jk}^{i}\dot{\delta}_{i}\otimes dx^{j}\otimes dx^{k},
\label{torsion_D}
\end{equation}%
where $R_{~jk}^{i}$ are the components of the curvature of the Ehresmann
connection $\overset{\alpha }{N}$.

\item The curvature of $D$ is:\textit{\ }%
\begin{eqnarray}
\mathbb{R} &=&R_{j~kl}^{~i}\delta _{i}\otimes dx^{j}\otimes dx^{k}\otimes
dx^{l}+~R_{j~kl}^{~i}\dot{\delta}_{i}\otimes \delta y^{j}\otimes
dx^{k}\otimes dx^{l}+  \label{curvature_D} \\
&&+B_{j~kl}^{~i}\delta _{i}\otimes dx^{j}\otimes dx^{k}\otimes \delta y^{l},
\notag
\end{eqnarray}%
where:%
\begin{equation}
\ R_{j~kl}^{~i}=\dfrac{1}{2}(E_{~k}^{i})_{\cdot jl},\ B_{j~kl}^{~i}=B_{\cdot
jkl}^{i}.  \label{curvature_components}
\end{equation}%
In particular, the Ricci tensor of $\overset{\alpha }{D}$ is given by the
Hessian with respect to the fiber coordinates of the trace $E_{~i}^{i}$: 
\begin{equation}
R_{jl}=-\dfrac{1}{2}(E_{~i}^{i})_{\cdot jl}=R_{j~li}^{~i}.
\label{Ricci tensor}
\end{equation}

\item Conversely, the tidal tensor $E$ of $\overset{\alpha }{N}$ and its
trace are obtained\footnote{%
Expression (\ref{relation_E_R}) points out an almost complete similarity
between the tidal tensor and the notion of flag curvature in Finsler
geometry. The difference consists in the metric tensor used in raising and
lowering indices, which is not the Finslerian one corresponding to $\overset{%
\alpha }{L}$ -- and which leads to somehow different properties.} in terms
of $\mathbb{R}$ as:%
\begin{equation}
E_{~k}^{i}=R_{j~kl}^{~i}y^{j}y^{l},~\ E_{~i}^{i}=-R_{jl}y^{j}y^{l}.
\label{relation_E_R}
\end{equation}%
In particular, for $\alpha =0,$ we have:\ 
\begin{equation*}
\overset{0}{E}\overset{}{_{~k}^{i}}=r_{j~kl}^{~i}y^{j}y^{l}=e_{~k}^{i},~\ 
\overset{0}{E}\overset{}{_{~i}^{i}}=e_{~i}^{i}.
\end{equation*}
\end{enumerate}

\section{Geodesics and geodesic deviations}

Consider a 1-parameter variation $c=c(t,\varepsilon ),$ $c(t,0)=x(t),$ of a
curve $t\mapsto (x^{i}(t))$ on $M.$ We denote by $V$ the complete lift of
the velocity $\dot{x}^{i}(t)\partial _{i}$ to $TM,$ expressed in the adapted
basis as:%
\begin{equation}
V:=y^{i}\delta _{i}+\dfrac{\delta y^{i}}{dt}\dot{\delta}_{i},~\ y^{i}=\dfrac{%
dx^{i}}{dt}  \label{V}
\end{equation}%
and by $\dfrac{D}{dt},$ the covariant derivative $D_{V}.$ The complete lift
of the deviation vector field $w$ along the curve $t\mapsto (x^{i}(t))$ is:%
\begin{equation*}
W=w^{i}\delta _{i}+\dfrac{\delta w^{i}}{dt}\dot{\delta}_{i}.
\end{equation*}

In the previous section, we have actually proved that, for extremal curves
of $\overset{\alpha }{L},$ the complete lift $V$ is horizontal (with respect
to $\overset{\alpha }{N})$:\ $V=hV.$ These curves can be also characterized
in terms of covariant derivatives attached to $\overset{\alpha }{D}.$

\begin{proposition}
\begin{enumerate}
\item Extremal curves for the Randers-type Lagrangian $\overset{\alpha }{L}$
obey the equation:%
\begin{equation}
D_{V}(hV)=0.  \label{geodesics_D}
\end{equation}

\item Deviations of geodesics (\ref{geodesics_D}) are given by:%
\begin{equation}
D_{V}^{2}(hW)=R(V,W)(hV);  \label{geodesic_deviation}
\end{equation}%
in local coordinates, this is:%
\begin{equation}
\dfrac{D^{2}w^{i}}{dt^{2}}=E_{~k}^{i}w^{k};  \label{geodesic_deviation_tidal}
\end{equation}%
here, all covariant derivatives are considered "with reference vector $y$", 
\cite{Shen}, i.e., in their local expressions, $G_{~j}^{i}=G_{~j}^{i}(x,y),$ 
$G_{~jk}^{i}=G_{~jk}^{i}(x,y).$
\end{enumerate}
\end{proposition}

\begin{proof}
1)\ Along the curve $t\mapsto (x^{i}(t)),$ we have:%
\begin{equation*}
D_{V}(hV)=\dfrac{D}{dt}(y^{i}\delta _{i})={\Large (}\dfrac{dy^{i}}{dt}%
+G_{~jk}^{i}y^{j}y^{k}{\Large )}\delta _{i}={\Large (}\dfrac{dy^{i}}{dt}%
+G_{~j}^{i}y^{j}{\Large )}\delta _{i}=\dfrac{\delta y^{i}}{dt}\delta _{i},
\end{equation*}%
where $y^{i}=\dot{x}^{i}.$ The extremality condition $\dfrac{\delta y^{i}}{dt%
}=0$ is therefore equivalent to $D_{V}(hV)=0.$

2)\ Differentiating (\ref{geodesics_D}) with respect to $W,$ we get:%
\begin{equation}
0=D_{W}D_{V}(hV)=D_{V}D_{W}(hV)+R(W,V)(hV)  \label{*}
\end{equation}%
(where we have taken into account that $[V,W]\equiv \lbrack \dfrac{\partial 
}{\partial t},\dfrac{\partial }{\partial \varepsilon }]=0$); further, $%
D_{W}(hV)=hD_{W}V=hD_{V}W+hT(W,V)=D_{V}(hW)$ (since $T(W,V)$ is vertical).
Thus, (\ref{*})\ becomes:%
\begin{equation*}
D_{V}^{2}(hW)+R(W,V)(hV)=0,
\end{equation*}%
which immediately yields (\ref{geodesic_deviation}). Expressing the term $%
R(V,W)(hV)$ in local coordinates and taking into account that $%
B_{~jkl}^{i}y^{l}=0,$ we get (\ref{geodesic_deviation_tidal}).
\end{proof}

\section{Basic equations of gravitational and electromagnetic fields}

In the following, we will apply the above construction to the case when $%
g_{ij}$ describes the gravitational field and $A=A_{i}dx^{i}$ in (\ref%
{randers}), is the 4-potential of the electromagnetic field. The
differential forms $A$ and $F$, \ref{comps_F_A}), can be lifted to
horizontal forms on $TM,$ which we will denote in the same manner. Unless
elsewhere specified, the parameter $\alpha \not=0$ is arbitrary.

Consider the \textit{angular metric, \cite{Shen}: }$h=h_{ij}dx^{i}\otimes
dx^{j}$ (regarded as a horizontal tensor on $TM$) given by:%
\begin{equation*}
h_{ij}=g_{ij}-l_{i}l_{j}.
\end{equation*}

The angular metric has the properties:%
\begin{equation}
h_{ij}=\left\Vert y\right\Vert l_{i\cdot j},~\ h_{ij}y^{j}=0.
\label{rel_h_l}
\end{equation}

\bigskip

The electromagnetic 2-form $F$ can be expressed in terms of the $\delta _{j}$%
-covariant derivatives of the 1-form $l=l_{i}dx^{i}.$ More precisely, we
have:%
\begin{equation}
D_{\delta _{j}}l_{i}=~\overset{0}{D}_{\delta _{j}}l_{i}-B_{~j}^{k}l_{i\cdot
k}-B_{~ij}^{k}l_{k}.  \label{F_derivs_D}
\end{equation}

It is easy to see that $\overset{0}{D}_{\delta _{j}}l_{i}=0.$ Evaluating the
remaining terms with the help of (\ref{expr_Bij}) and taking into account (%
\ref{rel_h_l}), we get that:%
\begin{equation}
D_{\delta _{j}}l_{i}=\dfrac{\alpha }{2}F_{ij};  \label{expr_F_1}
\end{equation}%
the above can be also written in the form (which reminds \cite{Ingarden}, 
\cite{Lagrange}, \cite{Miron-Buchner})

\begin{equation}
\alpha F_{ij}=D_{\delta _{j}}l_{i}-D_{\delta _{i}}l_{j},  \label{e-m. tensor}
\end{equation}%
in coordinate-free notation: $\alpha F=h(dl).$

We are now able to express Einstein-Maxwell equations in terms of tidal
tensors attached to $D=\overset{\alpha }{D},~\alpha \not=0$.

\textbf{A. Homogeneous Maxwell equations}\newline
\qquad Using (\ref{e-m. tensor}) and Ricci identities, \cite{Lagrange}, in
order to commute double covariant derivatives of $l,$ we obtain: 
\begin{equation}
-\alpha \left\Vert y\right\Vert (D_{\delta _{k}}F_{ij}+D_{\delta
_{j}}F_{ki}+D_{\delta _{i}}F_{jk})y^{k}=\tilde{E}_{[ij]},  \label{E1}
\end{equation}%
where:

\begin{equation}
\tilde{E}_{ij}=h_{ik}E_{~j}^{k}.\   \label{E0}
\end{equation}

Expressing $D=\overset{0}{D}+B,$ we notice that the terms involving the
contortion $B$ cancel each other, hence\ the above relation is actually:%
\begin{equation}
\tilde{E}_{[ij]}=-\alpha \left\Vert y\right\Vert (\overset{0}{D}_{\delta
_{k}}F_{ij}+\overset{0}{D}_{\delta _{j}}F_{ki}+\overset{0}{D}_{\delta
_{i}}F_{jk})y^{k}.  \label{E2}
\end{equation}%
Since $F_{ij}=F_{ij}(x)$ is projectable to $M,$ we can write (\ref{E2}) as:%
\begin{equation*}
\tilde{E}_{[ij]}=-\alpha \left\Vert y\right\Vert (\nabla _{\delta
_{k}}F_{ij}+\nabla _{\delta _{j}}F_{ki}+\nabla _{\delta _{i}}F_{jk})y^{k}.
\end{equation*}%
We have thus proven:

\begin{proposition}
Homogeneous Maxwell equations are written in terms of the tidal tensor $E$
as:%
\begin{equation}
\tilde{E}_{[ij]}=0,  \label{homogeneous_Max}
\end{equation}%
with $\tilde{E}$ as in (\ref{E0}).
\end{proposition}

\textbf{B. Inhomogeneous Maxwell equations}

Decomposing the curvature $R_{~jk}^{i}$ in terms of $r_{~jk}^{i}=~\overset{0}%
{R}\overset{}{_{~jk}^{i}}$ and $B_{~j}^{i}$ and contracting by $y^{k},$ we
get by direct computation that: 
\begin{equation}
E_{~i}^{i}=~e_{~i}^{i}-~\overset{0}{D}_{\delta
_{i}}(2B^{i})+B_{~i}^{l}B_{~l}^{i}.  \label{trace_E}
\end{equation}%
The derivative term is: $-\overset{0}{D}_{\delta _{i}}(2B^{i})=~\overset{0}{D%
}_{\delta _{i}}(\alpha \left\Vert y\right\Vert F_{~j}^{i}y^{j})=\alpha
\left\Vert y\right\Vert y^{j}\overset{0}{D}_{\delta _{i}}F_{~j}^{i}=\alpha
\left\Vert y\right\Vert y^{j}\nabla _{\partial _{i}}F_{~j}^{i}.$

As a consequence, we get:

\begin{proposition}
Inhomogeneous Maxwell equations \textit{are expressed in terms of tidal
tensors as:}%
\begin{equation}
E_{~i}^{i}=~e_{~i}^{i}-4\pi \alpha \rho _{c}\left\Vert y\right\Vert
^{2}+B_{~i}^{l}B_{~l}^{i},  \label{inhom_maxwell_tidal}
\end{equation}%
where $\rho _{c}=-J^{i}l_{i}$ and $~e_{~i}^{i}=\overset{0}{E}\overset{}{%
_{~i}^{i}}.$
\end{proposition}

\begin{remark}
\begin{enumerate}
\item An alternative expression can be obtained if we notice that $B^{i}=%
\dfrac{1}{2}g^{ih}\delta _{h}(\left\Vert y\right\Vert ^{2})=:\dfrac{1}{2}%
\delta ^{i}\left\Vert y\right\Vert ^{2};$ we get: $\overset{0}{D}_{\delta
_{i}}B^{i}-B_{~i}^{h}B_{~h}^{i}=D_{\delta _{i}}B^{i}=\dfrac{1}{2}D_{\delta
_{i}}(\delta ^{i}\left\Vert y\right\Vert ^{2}),$ i.e., we have the equality:%
\begin{equation}
E_{~i}^{i}=~e_{~i}^{i}-2\pi \alpha \rho _{c}\left\Vert y\right\Vert ^{2}-%
\dfrac{1}{2}D_{\delta _{i}}(\delta ^{i}\left\Vert y\right\Vert ^{2}).
\label{expression_E_ii}
\end{equation}

\item The trace of $\tilde{E}$ is: $\tilde{E}_{~i}^{i}=g^{ij}\tilde{E}%
_{ji}=g^{ij}h_{~j}^{k}E_{ki}=g^{ij}(\delta
_{j}^{k}-l^{k}l_{j})E_{ki}=E_{~i}^{i}-l^{k}l^{i}E_{ki}.$
\end{enumerate}
\end{remark}

Taking into account relation (\ref{relation_E_R}), we obtain that $%
l^{k}l^{i}E_{ki}=0$, which means that $\tilde{E}_{~i}^{i}=E_{~i}^{i}.$ In
other words, we can use in (\ref{inhom_maxwell_tidal}) either of the
versions $\tilde{E}_{~i}^{i}$ or $E_{~i}^{i}.$

\textbf{C. Einstein field equations}

Einstein field equations can also be obtained from the trace $E_{~i}^{i},$ (%
\ref{trace_E}), if we substitute, this time, the term $e_{~i}^{i}=\overset{0}%
{E}\overset{}{_{~i}^{i}}.$ Contracted with $y^{i}y^{j},$ the Einstein field
equations $r_{ij}=8\pi (T_{ij}-\dfrac{1}{2}T_{~l}^{l}g_{ij})$ become: 
\begin{equation}
e_{~i}^{i}=-8\pi (T_{ij}y^{i}y^{j}-\dfrac{1}{2}T_{~l}^{l}\left\Vert
y\right\Vert ^{2}).  \label{efe_tidal}
\end{equation}

The stress energy tensor $T_{ij}$ can be decomposed as: $T_{ij}=\overset{em}{%
T}_{ij}+\overset{m}{T},$ where $\overset{em}{T}_{ij}$is the stess-energy
tensor of the electromagnetic field, \cite{Landau}:%
\begin{equation*}
\overset{em}{T}_{ij}=\dfrac{1}{4\pi }(-F_{~i}^{h}F_{hj}+\dfrac{1}{4}%
F^{kh}F_{kh}g_{ij})
\end{equation*}%
and $\overset{m}{T}_{ij}$ is the stress-energy tensor of matter (and/or
other fields). The electromagnetic stress-energy tensor $\overset{em}{T}%
_{ij} $ has zero trace $\overset{em}{T}\overset{}{_{~l}^{l}}=0$, hence, in (%
\ref{efe_tidal}), $T_{~l}^{l}=\overset{m}{T}\overset{}{_{~l}^{l}}.$

On the other side, taking in (\ref{expr_F_1}) covariant derivative by $%
\delta _{k}$ and succesively contracting by $g^{jk}$ and $\left\Vert
y\right\Vert ^{2}l^{i},$ we obtain the identity:

\begin{equation*}
\left\Vert y\right\Vert ^{2}l^{i}\square l_{i}=\overset{0}{D}_{\delta
_{i}}B^{i}+\alpha ^{2}(-F^{h}F_{h}+\dfrac{1}{4}F^{kh}F_{kh}\left\Vert
y\right\Vert ^{2}),
\end{equation*}%
where $\square l_{i}:=g^{jk}D_{\delta _{k}}D_{\delta _{j}}l_{i}.$ The latter
equation is actually: $4\pi \alpha ^{2}\overset{em}{T}_{ij}y^{i}y^{j}=\left%
\Vert y\right\Vert ^{2}l^{i}\square l_{i}-\overset{0}{D}_{\delta _{i}}B^{i}.$

Consequently, relation (\ref{efe_tidal}) is equivalent to:%
\begin{equation}
e_{~i}^{i}=-\dfrac{2}{\alpha ^{2}}(\left\Vert y\right\Vert ^{2}l^{i}\square
l_{i}-\overset{0}{D}_{\delta _{i}}B^{i})-8\pi (\overset{m}{T}_{ij}y^{i}y^{j}-%
\dfrac{1}{2}\overset{m}{T}\overset{}{_{~l}^{l}}\left\Vert y\right\Vert ^{2}),
\end{equation}

Replacing into the expression (\ref{trace_E}) of $E_{~i}^{i}$ and denoting $%
\rho _{m}:=\overset{m}{T}_{ij}l^{i}l^{j},$ we finally have:%
\begin{equation}
\dfrac{1}{\left\Vert y\right\Vert ^{2}}E_{~i}^{i}=\dfrac{2}{\left\Vert
y\right\Vert ^{2}}\{(\dfrac{1}{\alpha ^{2}}-1)\overset{0}{D}_{\delta
_{i}}B^{i}+B_{~l}^{i}B_{~i}^{l}\}-\dfrac{2}{\alpha ^{2}}l^{i}\square
l_{i}-8\pi (\rho _{m}-\dfrac{1}{2}\overset{m}{T}\overset{}{_{~l}^{l}}\}.
\label{efe_tidal1}
\end{equation}

\bigskip 

\textbf{D. Equations of motion of charged particles}\newline
\qquad Equations of motion of a charged particle,\cite{Landau}, are nothing
but (\ref{Lorentz_eq}):%
\begin{equation}
\dfrac{\overset{\alpha }{D}y^{i}}{dt}=0,~\ y=\dot{x},  \label{Lorentz_eqns}
\end{equation}%
in which, this time, we set: 
\begin{equation}
\alpha =\dfrac{q}{m}.  \label{alpha_Lorentz}
\end{equation}%
For particles having the same ratio $\dfrac{q}{m},$ worldline deviation
equations are given by 
\begin{equation}
\dfrac{\overset{\alpha }{D}\overset{}{^{2}}w^{i}}{dt^{2}}=E_{~j}^{i}w^{j},~\
\alpha =\dfrac{q}{m}.  \label{worldline_deviation}
\end{equation}

\section{Particular cases}

\textbf{A. Gravity only: }

In this case, we have $B^{i}=0,$ which means that all the affine connections 
$\overset{\alpha }{D},$ \ $\alpha \in \mathbb{R},$ have vanishing contortion
(they actually coincide)\ and $R_{j~kl}^{~i}=r_{j~kl}^{~i}$. The tidal
tensor is given by $e_{~j}^{i}=r_{l~jk}^{~i}y^{l}y^{k}$ and equations (\ref%
{Costa_gravi}) become:%
\begin{equation}
\begin{array}{c}
r_{ij}=8\pi (\overset{m}{T}_{ij}-\dfrac{1}{2}g_{ij}\overset{m}{T}\overset{}{%
_{~l}^{l}}) \\ 
r_{jk}=r_{kj}%
\end{array}%
~\ \ \ \ \ \ 
\begin{array}{c}
\Leftrightarrow \\ 
\Leftrightarrow%
\end{array}%
~\ \ \ \ \ 
\begin{array}{c}
\dfrac{1}{\left\Vert y\right\Vert ^{2}}e_{~i}^{i}=-4\pi (2\rho
_{m}-T_{~i}^{i}) \\ 
\tilde{e}_{[ij]}=0,%
\end{array}
\label{gravi_only_tidal}
\end{equation}%
where, this time, $\rho _{m}=T_{ij}l^{i}l^{j}$.

\textbf{B. Electromagnetism in flat Minkowski space}

In this case, we have $\gamma _{~jk}^{i}=0,$ $e_{~j}^{i}=0$ and $%
G_{~jk}^{i}=B_{~jk}^{i}$. In the expression (\ref{curvature_D}) of the
curvature of $\overset{\alpha }{D}$ ($\alpha \not=0$), the components $%
R_{j~kl}^{~i}$, $B_{j~kl}^{~i}$ only depend on $B$. 

Maxwell equations are written in terms of tidal tensors as:%
\begin{equation}
\ 
\begin{array}{c}
\nabla _{\partial _{i}}F^{ij}=4\pi J^{i} \\ 
\nabla _{\partial _{i}}F_{jk}+\nabla _{\partial _{k}}F_{ij}+\nabla
_{\partial _{j}}F_{ki}=0%
\end{array}%
~\ \ 
\begin{array}{c}
\Leftrightarrow  \\ 
\Leftrightarrow 
\end{array}%
~\ \ \ 
\begin{array}{c}
\dfrac{1}{\left\Vert y\right\Vert ^{2}}E_{~i}^{i}=-4\pi \alpha \rho _{c}+%
\dfrac{1}{\left\Vert y\right\Vert ^{2}}B_{~h}^{i}B_{~i}^{h} \\ 
\tilde{E}_{[ij]}=0.%
\end{array}%
.  \label{maxwell_tidal}
\end{equation}

Thus, we found analogous equations to the ones determined by Costa and
Herdeiro, \cite{Costa}, \cite{Costa2}, without resorting to any restriction
upon the derivatives of the deviation vector $w$.

\textbf{Acknowledgment. }The work was supported by the Sectorial Operational
Program Human Resources Development (SOP HRD), financed from the European
Social Fund and by Romanian Government under the Project number
POSDRU/89/1.5/S/59323.

\end{document}